\documentclass[12pt,preprint]{aastex}

\newcommand \radm{rad/m$^2$}
\newcommand \be{\begin{equation}}
\newcommand \ee{\end{equation}}
\newcommand \etal{{\it et al.\/}}
\newcommand \eg{{\it e.g.}}
\newcommand \cf{{\it cf.}}

\newcommand \ltw{\>\hbox{\lower.25em\hbox{$\buildrel <\over\sim$}}\>}
\newcommand \gtw{\>\hbox{\lower.25em\hbox{$\buildrel >\over\sim$}}\>}
\newcommand \Bparmean{\langle B_{\parallel} \rangle}

\shorttitle{Magnetic Fields in Cluster Cores}
\shortauthors{Eilek \& Owen}

\begin{document}

\title{Magnetic Fields in Cluster Cores:  Faraday Rotation in A400 and
A2634}

\author{ Jean A. Eilek}

\affil{Physics Department, New Mexico Tech \\
Socorro NM 87801}

\and

\author{ Frazer N. Owen}
\affil{National Radio Astronomy
Observatory\footnote{The National Radio
Astronomy Observatory is operated by  Associated Universities, Inc., under
a cooperative agreement with the National Science Foundation.}, \\
Socorro, New Mexico  87801
}

\begin{abstract}

 We present Faraday rotation data for radio sources
in the centers of the Abell clusters A400 and A2634.  These clusters
contain large ($\gtw 100$kpc), tailed radio sources, each attached to the
central cD galaxy.  These clusters do not have strong
cooling cores. Our data extend
previous work on rotation measure in cluster centers to larger scales
and non-cooling clusters.  The rotation measure, and thus the
magnetic field, is ordered on scales $\sim 10-20$ kpc in both
clusters.  The geometry of the rotation measure appears to be
determined by the distribution of the X-ray emitting gas, rather than
by the radio tails themselves.  We combine our data with previously
published X-ray and radio data in order to analyze the magnetic fields
in all 12 clusters whose central radio sources have been imaged
in rotation 
measure. We find that the fields are dynamically significant in most
clusters.  We argue that the Faraday data measure fields in the
intracluster medium, rather than in a skin of the radio source.
Finally, we consider the nature and maintenance of the magnetic
fields in these clusters, and conclude that either the cluster-wide
field exists at similar levels, or that a weaker cluster-wide field is
amplified by effects in the core.

\end{abstract}

\keywords{galaxies:  clusters -- galaxies:  magnetic fields --
magnetic fields}

\section{Introduction}

The physical state of the plasma in clusters of galaxies is far from
understood.  One important question is whether that plasma is
magnetized, and if so at what levels.  Some important steps toward
answering this question have already been made. 

There is some evidence that modest fields exist throughout
the intracluster medium (ICM).  
One approach uses the integrated rotation measure (RM) of
sources in or behind clusters. Hennessy \etal\  (1989) compared
sources seen through cluster cores to a non-cluster control sample,
and found no significant excess RM in the cluster sample.  In a
similar study, Kim \etal\ (1991) did find excess RM in sources in and
behind clusters (although the effect was less strong when they
excluded radio sources embedded in the clusters).  Clarke, Kronberg \&
B\"ohringer (2001; CKB)  carried out a similar study with a more
uniform sample, also using both background and cluster-member sources.
They also find 
evidence for an intrinsic cluster field.  The excess RM values
suggested by the Kim \etal\ and the CKB work are  consistent with
modest magnetic fields, at a few  $\mu$G levels, throughout the
cluster.  This work is not yet conclusive, however.  Unresolved data 
can underestimate the true RM of a particular source.  In addition,
neither the Kim \etal\ nor the CKB work can distinguish between a
few $\mu$G field intrinsic to the cluster, and a much weaker field
which is enhanced only around the observed radio galaxies.

Another approach relies on the diffuse synchrotron radio haloes have
been detected in several clusters (\eg, Feretti \& Giovannini 1996).
Ongoing work (\eg, Owen, Morrison \& Voges 1999, Feretti 1999)
suggests these are  not as rare as has been thought.  The simplest
minimum-pressure or equipartion analyses find the field associated
with these haloes is  $\sim 0.1 - 1 \mu$G.  X-ray   
spectra  from these halo clusters also allow the possibility of
detecting microwave background photons which have been inverse Compton
scattered up to the X-ray range.  A non-detection in the X-rays leads
to a lower limit to the magnetic field, generally in the
sub-microgauss range (\eg, Henriksen, 1998).  Reported detections, if
confirmed, also suggest similar field values (\eg, Fusco-Femiano
\etal\ 1999).  

Thus, both the RM data and the radio halo data suggest magnetic fields
in the $\sim \mu$G range,   throughout
the volume of the cluster.  These  analyses have  
been carried out assuming a uniform, volume-filling magnetic field.
We note that the  data are also consistent with higher fields if the
filling factor is small (which we argue below is likely to be the case). 

Additional information comes from resolved RM images of radio sources
within the cluster.  Sources in cluster center are easiest to 
interpret in terms of the associated magnetic field, due to the
certainty of their position within the ICM.  The best-studied clusters
so far are those with strong central X-ray peaks, associated with
cooler gas.  These are the clusters that have been interpreted as
``cooling flows''.  Several high-quality images exist for the  
RM associated with radio sources in the
centers of such clusters  (\eg, Ge \& Owen 1994; Taylor,
Allen \& Fabian 1999; and references therein).  These authors find high RM
values,  several thousand \radm, with order  scales $\sim 1 -  10$ kpc
are  typical of radio sources in this environment. 

Thus, present evidence supports a modest magnetic field, at 
microgauss levels, throughout most clusters.  In addition, a much
stronger field exists in the inner region of strong cooling cores.
The connection between these two results is not yet clear.  
In this paper we extend the data base by presenting RM images of two
larger, tailed radio galaxies in clusters which do not have strong
cooling cores.  We consider 3C75 in Abell 400, and 3C465 in Abell
2634.  Both of these clusters have only weak central cooling.  Both of
the radio sources extend to $\gtw$ 100 kpc, so that we can study a
more significant fraction of the cluster core.  We find the RM is
ordered, on scales $\sim 10 - 20$ kpc, and that the order seems
to depend more on the geometry of the X-ray gas than of the radio
tails themselves.  By combining this with 
gas densities determined from X-ray deprojection, we find the magnetic
fields are likely to be dynamically important in these two clusters.

We begin by introducing the two clusters  and presenting our new data
(in \S 2).  After reviewing basic assumptions necessary for RM
analysis (in \S 3), in \S 4 we  place our results in the context of
all nearby clusters for  which  high-quality RM data of the central
source is available.  In \S 5 we consider possible mechanisms by which
the magnetic field can be maintained.  We present our conclusions in
\S 6.

\section{The Data:  X-Ray Analysis and Rotation Measure Imaging}

\subsection{Abell 400}

 Abell 400 is a richness class 1,  Rood-Sastry I cluster.  Its dynamics
have been analyzed by Beers \etal\ (1992).  They find a mean redshift
$z = .0235$, which corresponds to 26.5 kpc/amin, using ${\rm H}_o = 75$
km/s-Mpc.   After separating out a background group of galaxies to the
NE,   they find that the cluster itself has substructure.  The
central, elliptical-rich population is composed of two subgroups,
separated by $\sim 700$ 
km/s.  These groups are not apparent in the galaxy distribution 
projected on the sky.  From this, they conclude that a merger is
taking place, close to the line of sight.  
The cluster centers on a dominant galaxy with  two nuclei separated by
435 km/s (\eg, Balcells \etal\ 1995).  The double nuclei have been 
interpreted as due to projection (Tonry 1985), as closely interacting
(Lauer 1988) and as merging clump centers (Beers \etal, also Balcells
\etal).    

The X-ray image is fairly smooth, with some elongation of the inner
isophotes  (see also Beers \etal\  1992 who present ellipse fits to the
inner isophotes). Figure 1 shows the image from the ROSAT PSPC,
together with the central radio galaxy.  Abell 400 does not have a
strong cooling core.  White \etal 
(1997) agree with earlier authors in finding that any spherical mass
inflow is small, $\dot M \ltw 10 M_{\sun}$/yr.  

The central galaxy is host to an unusual radio source:  
each nucleus gives rise to a radio jet, and the 
jets merge into the striking tailed radio source, 3C75 (Owen \etal\ 
1985).   Figure 1 shows that the tails can be detected to $\sim 7$
amin $\sim 185$ kpc at 327 MHz.  They bend to the NE, away from the
X-ray centroid and axis of elongation.  Note  that our RM data (based
on higher frequencies than the image in Figure 1) extend only to
$\sim 200$ asec $\sim 90$ kpc. 

\subsection{Abell 2634}

Abell 2634 is  a richness class 1, Rood-Sastry cD cluster. It has
a mean $z = .0313$ (Pinkney \etal\  1993), which corresponds to 35.2
kpc/amin.  Its 
dynamics and structure have been studied by Pinkney \etal\ and
by Scodeggio \etal\ (1995).  While the dynamical situation is not totally
clear, both papers agree on the most likely picture.  The cluster
itself does not show any evidence for substructure, either in velocity
or physical space.  The cluster sits in a complex area of the sky which
also contains the lower redshift 
Pisces-Perseus supercluster.  Two groups of galaxies sit $\sim
1^{\circ}$ to the east, one high-velocity and low-velocity relative to
A2634.  When these groups are excluded, the velocity and spatial
structure of A2634 are consistent with a relaxed cluster.  Thus based on
the velocity field there is no evidence
that A2634 is currently undergoing a major merger.  

The X-ray image of this cluster is more complex than the optical data
would suggest.  On large scales, the isophotes are more or less
regular;  but on small scales a striking central elongation appears. 
Figure 2 shows the image from the ROSAT PSPC, together with the
central radio galaxy (Eilek \etal\  1984; also {\it c.f.} 
Schindler \& Prieto 1996 who present inner X-ray isophotes).  The
distorted central X-ray distribution 
has no counterpart in the optical image.  Although Schindler \&
Prieto (1996) argue against this distortion being caused by a merger,
Roettiger,  Loken  \& Burns (1997) find that such structures can indeed
occur in 
the later stages of a merger.  They find density elongations and bulk
flows along the merger axis. As with Abell 400, there is no
significant evidence of a cooling core in A2634 (White \etal\ 1997, who
agree with earlier work).  

The central galaxy also has a double nucleus (Lauer 1988).  One of the
nuclei gives rise to a large tailed radio 
galaxy, 3C465 (Eilek \etal\ 1984).  The radio tails can be detected to
$\sim 7$ amin $\sim 250$ kpc at 327 MHz.  They bend to the SW, towards
(and around) the X-ray elongation. Our RM map, again based on high
frequencies,  covers only the inner region, extending to $\sim 230$
asec $\sim 135$ kpc.        

\subsection{X-ray Analysis}

The two clusters were observed in pointed observations with the
ROSAT PSPC.   We carried out a simple
deprojection analysis to determine the density and
temperature of the inner regions of each cluster.  As the inner part of
each cluster is distorted from circular symmetry, we allowed for this
in our deprojection.   (Beers \etal\ 1992 present ellipse fits to the
surface brightness for A400, as do Schindler \& Prieto 1996 for
A2634.  These papers display the elongated isophotes for each
cluster, which can also be seen in Figures 1 and 2.)  We used IRAF to
fit elliptical contours to the surface  
brightness of each cluster, and to determine the flux from each
elliptical annulus.  The effective scale of the ellipse is
$\sqrt{ab}$, if $a$ and $b$ are the semi-major and semi-minor axes.
In order to determine the emitting volume of a particular ellipse, we
assumed the major axis of each ellipse  lies in the plane of the sky,
and that the volumes were prolate ellipses.   We then used an
``onion-skin'' method to deproject the data, as follows.  At a given
point on the sky, say 
some effective distance $R$ from the X-ray peak, the surface brightness is a
linear sum of the contributions from all elliptical shells lying at or
outside $R$.  If the number of shells is finite, as is the
case with discretely sampled data, then the surface brightness can be
deprojected using simple linear analysis to determine the intensity in
each elliptical volume.  (Fabian \etal\ 1981 describe a similar method,
however assuming spherical symmetry). 

We separated the X-ray data into the four R4 to R7 bands covering
0.5 to 2.05 keV within
the ROSAT window (Snowden et al 1994), and carried out
our analysis separately on each band.  This gave us a four-point spectrum
for each shell. Using XSPEC to estimate the emissivity as a function of
temperature in each band, while holding metallicity
constant at 0.3 solar which is typical of similar clusters (e.g. Edge \&
Stewart 1991), we fit both temperature and density in each shell.

Table 1 shows the density and temperature we derive for each
elliptical shell.  The density values are consistent with other
analyses of non-cooling clusters such as these, and also with the
analysis of Eilek \& Markovic (2001).  They find the 
size of the ``gas core'' (at which the density drops to 20\% of its
central value) to be $\sim 300$ kpc for A400, $\sim 400$ kpc for
A2634. The temperature structure from Table 1 is worth comment.  We
find that both clusters have cooler gas in the inner $\sim 25 - 20$
kpc, outside of which the temperature rises to a warmer value which
describes the cluster as a whole.  Schindler \& Prieto find a similar
result for the core of A2634;  we are not aware of any spectral
analysis for the core of A400.  Our results for the large-scale
temperatures are a bit lower than other authors find.  White \etal\
(1997) find 2.1 keV for A400 and 3.4 keV for A2634 (full clusters),
while Fukazawa \etal\ (1998) find 2.3 keV in A400, and 3.7 keV in
A2634, for the region outside of 100 kpc.  The reason for this
discrepancy is not clear;  we suspect that our use of elliptical
isophotes combines cool and hot gas in a different way than the
circular rings used by other authors.  In addition, the poor
temperature sensitivity of ROSAT to high-temperature gas may
contribute to the discrepancy. 

For the purposes of this paper, we use this analysis to extract
typical densities and pressures characteristic for the region in which
we have Faraday rotation data,  namely,  the inner 100 - 200
kpc of each cluster.  This will be used in \S 4, below.

\subsection{Rotation Measure Imaging}

VLA observations were carried out using the C and D arrays in
four frequency bands each within the C-band (4.5-5.0 GHz) and the
X-band (8.0-8.9 GHz) for a total of eight frequencies. Standard
calibration, including instrumental polarization, was applied to the
data. After cleaning and imaging  using AIPS, we determined the
rotation measures using a modified version 
of the AIPS task, RM, which took all eight frequencies as input. The RM
program also produces an estimate of the intrinsic magnetic field direction
by correcting the polarization vectors to zero wavelength from the rotation
measure and adding 90 degrees.

We thus determined the  rotation measure, and
also the intrinsic direction of the magnetic field,
projected on the plane of the sky. In Figures 3 and 4 we
show the projected magnetic field directions 
for each source.   We see that the two sources have different
characteristics in projected field directions.  3C465, in A2634, has
a projected field which more or less follows the radio tails.  This is
consistent with other sources in which the magnetic field agrees 
with the apparent flow direction.  However, 3C75 in A400 is more
complex;  interesting small-scale loops and other field structures are
apparent in Figure 3.  We suspect this is due to the more complex flow
field of this source (recalling that the total intensity image, Owen
\etal\ 1985, shows loops and twists which suggest the two jets are
interacting strongly).  As our goal in this paper is the RM
information, we defer analysis of the polarization in these two
sources to a later paper.

We present the RM data in several ways.  In Figures 5 and 6 we present
RM images for each object, with the colors chosen to highlight
positive and negative RM values.  Alternating positive and negative RM
patches are apparent in each source;  the RM in each case appears to
have a characteristic ``order scale'' of 10 -20 kpc.  These patches
are well resolved; our restored beam was 3x3 asec ($\sim 1.3$ kpc for
3C75, $\sim 2.2$ kpc for 3C465). We can characterize the RM
distribution in terms of a mean and an rms.  For A400, the mean is -7.6
\radm, with rms 100 \radm.  For A2634, the mean is -25 \radm, and
the rms is 120 \radm.   The non-zero means could
be due to galactic foreground, they could be a small mean RM due to the
cluster itself, or they could simply be statistical fluctuations
either in RM's of the distant radio sources or structure in the
galactic RM distribution.   We cannot distinguish between these  options,
but this is not critical for our analysis here. 

Structure in the RM can be studied with other displays.  In Figures 7
and 8, we show the absolute value of the RM as a function of distance
from the galactic core.  These  data are means within circular apertures,
centered on the galactic core; the horiozontal axis is the radius of
each ring.  The mean magnitude of the RM is approximately constant over
$\sim 50$ kpc in A400, and drops by a factor $\sim 2$ over 80 kpc in
A2634.  We also find small patches of large RM magnitudes.  
Figures 9 and 10, we present color images of the absolute value of the
RM for  each source; the scaling is chosen to highlight the small
patches of extreme $|{\rm RM} |$ in the east tail of 3C465, and the west tail
of 3C75. 

\subsubsection{Rotation Measure Structure in Each Cluster} 

These data show that the overall properties of the RM distribution
in each cluster are quite similar. The images give a strong impression
of order in the RM, rather than being purely random. Each source
shows both positive 
and negative RM patches, with comparable areas showing positive and
negative values.  The projected scale of the patches is $\sim 10 - 20$
kpc in each object.  
What can we infer about the three-dimensional structure, or the
physical location, of the field causing the RM, in each object?  In
A400, one's first impression is 
of a set of positive and negative patches, without particular order.
On closer inspection, however, the two highly negative patches (RM
$\sim -250$ to -300 \radm) on the west tail stand out.  In addition, two
strongly positive patches (RM $\sim 150$ to 200 \radm) can be seen in
the east tail.  None of these areas correlate with any particular
feature in the radio tails.  It is tempting to interpret the two
positive areas and two negative areas as being continuous bands of
extreme RM, which happen to be projected on the radio tails.  With
this interpretation, it is interesting to note that the bands lie
roughly parallel to the major axis of the X-ray elongation (as shown
in Figure 1).

In A2634, order in the RM image is more striking.  Clear positive and
negative RM bands alternate along each tail.  However, the relation of
the RM to the tail geometry is different for the two tails.  Along
the west tail, the RM bands tend to lie across the tail; along the
east tail they tend to lie along it.  The extreme RM values here are
found in the east tail.  The radio hot spot shows strongly positive
and negative values, $\sim \pm 200$ \radm, alternating across the hot
spot.  This is the one place in the two sources where extreme RM
features appear to correlate with particular radio features.  The RM
bands just below this feature also have extreme RM values, in about
the same range.  However, they 
do not correlate with any particular radio feature.  Once
again, it is possible to interpret the data as continuous positive and
negative RM features, which happen to be projected onto the radio
tails.  In this case, however, the RM bands lie roughly perpendicular
to the major axis of the X-ray elongation (as shown in Figure 2).

\subsubsection{Location of the Faraday Screen}

The observed rotation measure is almost certainly due to an
intervening magnetized plasma, rather than the radio source itself.  
This can be seen from two lines of argument.

First, consider the alternative case,  in which the RM arises from the
radio source itself.  We can refer to Burn (1966),  also Laing (1984), to
understand the effects of internal Faraday rotation.  Internal RM will
depolarize;  by the time the electric vector has rotated by one
radian, the fractional polarization of the signal drops nearly to
zero.  In addition, internal RM causes the rotation angle $\chi$ to
deviate from the usual $\lambda^2$ behavior in all geometries except
the very simplest one, a resolved slab with fully uniform magnetic
field direction.  For $\chi \ltw 1$, the deviations are small
but measurable;  for $\chi > 1$, the deviations are significant.  The
observed polarization angle obeys $\chi \propto \lambda^2$ very well
in both sources.  An RM of 100 \radm\ leads to $\chi
\sim 1/3$ radian at 6 cm;  one might argue that deviations of a real
source from an ideal $\lambda^2$ law are small enough in this regime
not to be detected.  The same RM leads to $\chi \sim 3$
radian at 20 cm, which would depolarize the sources when observed at
wavelength.   This is not borne out by observations. Eilek \etal\
(1984) measured $\sim 20$\% fractional polarization at 20 cm. 
We also have unpublished 20 cm data for 3C75 which shows similarly
high polarization. Thus, it is quantitatively
difficult to ascribe the RM 
in these two sources to internal effects.

We  can also make two qualitative arguments.  First, many
other cluster-center radio galaxies have very high measured RM values
(summarized in \S 4).  All of these have high enough RM that the
observed polarization and rotation must come from a foreground screen;
internal effects as described above would depolarize the source and
cause strong deviations from the $\lambda^2$ law.  These clusters thus
have a magnetized ICM which provides the Faraday screen.  It seems
simplest also to ascribe the RM in our two sources to Faraday effects
from the magnetized ICM.  Second, one can compare the de-rotated magnetic
field vectors, in Figures 3 and 4, with the RM images, Figures 5 and
6.  There is little relation in either source between the field
direction, and the RM values.  If the RM were internal, we would expect
the geometry of the RM image to be reflected in the geometry of the
projected field.  This is not the case.  Thus, we conclude that the
intervening magnetized plasma which causes the RM is almost 
certainly the ICM in the two clusters.  

\section{Analysis:  Magnetic Fields in Cluster Centers}

The data tell us that a magnetic field exists in the cores of these
two clusters.  In this section we consider likely field models and
the magnetic field strength that each requires. 

\subsection{Basic Tools}

We first recall the basic expression for rotation measure.  Let $\chi$
be the angle of the electric vector, and $\lambda$ the observing
wavelength.  The fundamental relation is,
\be
{\rm RM} = { d \chi \over d \lambda^2} = { e^3 \over 2 \pi m^2 c^4} \int n
\mathbf{ B} \cdot d \mathbf{ l} ~~ {\rm rad/m^2} 
\ee
Note that we cannot measure the full vector magnetic field,
$\mathbf{ B}$. Rather, we
measure the density-weighted mean of its component along the line of
sight.  Expressing this numerically, and assuming a constant density
over length $L_{RM}$, we have
\be
{\rm RM} \simeq 810 n L_{RM} \Bparmean ~~ {\rm rad/m^2}
\ee
if $L_{RM}$ is the typical length scale of the magnetized plasma along
the line of sight, and $\Bparmean$ is the weighted algebraic mean of
the  component of $\mathbf{B} $, also along the line of sight. 
In this expression, $B$ is in $\mu$G, $n$ in cm$^{-3}$ and $L$ in kpc.  
 The importance of  
this line-of-sight averging is that, given a density and a scale
length, we are measuring only a lower limit to the magnetic field.
The field is 
likely to have components perpendicular to the line of sight (which
increases $B$ by $\sqrt{3}$, on average, from $\Bparmean$).  In 
addition, any sign reversals along the line of sight also lead to a
true $B_{\parallel}$ which is larger than $\Bparmean$.

Knowledge of the gas density is needed to interpret the RM data.
X-ray data reveal the distribution of thermal gas in the cluster.   
Some clusters have strong X-ray peaks, in which the high gas density
makes cooling important;  we call these cooling cores (CC's).  Other
clusters have flatter central X-ray profiles, in which 
cooling is not important;  we call these non-cooling clusters (NC's).  
Both cooling and non-cooling clusters show a gaseous core, of roughly
constant density, and an extensive outer region of declining gas
density.  The two types of clusters differ only in the size
and central density of the core.  Eilek \& Markovic (2001) find that
hydrostatic gas in a Navarro, Frenk \& White (1997)
potential describes the X-ray surface brightness well in 
both types of
clusters.  They give a Hubble-type approximation for the gas density, 
\be
n_g(r) \simeq { n_o \over (1 + r/a)^x}
\ee
where the
exponent $x \simeq 2.0$ for
CC's and  $x \simeq 2.7$ for NC's. Fits to the X-ray data show that
the core radii are much smaller for CC's than for NC's: $a \sim 100$
kpc for CC's, while $a \sim 600$ kpc for NC's.  The radius of the
``gas core'', defined as the region 
within which the gas density $\gtw 0.2 n_o$, is $\sim 90$ kpc for
CC's, and $\sim 400$ kpc for NC's.

\subsection{Possible Extent of the Magnetic Field}

What is the spatial extent of the magnetic field which causes the RM?
There are three possibilities.  (1)  The field
 may originate in the radio source
itself, and be mixed with the ambient gas in a ``skin'' of the radio
source.  (2)  The field may be a cluster-wide field,  and the radio
source simply provides a backlight to illuminate it. (3)  The
field may originate in a cluster-wide field, but be enhanced in the
center either by a cooling-driven inflow, or 
by interactions of the radio source with the ambient medium. 
We discuss each in turn.  We base our discussion on A400 and A2634,
and on the full set of well-imaged cluster-center sources (reviewed in
\S 4, below). 

\subsubsection{Fields in a skin of the radio source}

Shortly after the first detections of high RM, 
Bicknell \etal\  (1990) argued that it is solely due to the radio
galaxy.    They proposed that surface turbulence from a 
Kelvin-Helmholtz instability  
allows diffusion of particles and fields from the low-density,
radio-loud plasma into the denser  ICM.  The RM comes from
this mixing layer. This model does not require that
the cluster plasma contain any magnetic field. 
Other authors (Taylor \& Perley 1993, Ge \& Owen 1993)
argued against this idea.  They argued that this picture is
quantitatively unlikely. The high RM in smaller sources requires
either a mixing layer which is thick compared to the size of the source
(unlikely on physical grounds), 
or a strong  magnetic field in a thin mixing layer (which
would be overpressured and short-lived). Taylor \& Perley also 
point out that the RM in Hyd A is strong along the jet as well as
lobes (unlikely in an instability-driven model).  
We also find the mixing-layer model unlikely.   The RM distribution in
the two sources presented in this paper seems to be influenced more by
the elongation axis of the X-ray luminous gas than the direction of
the tails.  This is apparent by comparing Figures 1 and 5, or 2 and
6. This suggests to us a cluster origin.  

We also note geometrical arguments against the thin-skin model.  
(1) More cluster-center sources show asymmetric RM
patterns than symmetric ones: the RM distribution and
magnitude are quite different in the two lobes.  As one might expect
fluid instabilities to develop symmetrically in a symmetric radio
source, this argues against a mixing-layer origin. (2) One 
might expect edge effects from this mixing layer:  the RM image should
be limb-brightened, or perhaps limb-darkened.  We find no sign of limb
effects in the RM maps of A400, A2634, or any of the other
cluster-center sources.  (3) 
There are strong suggestions of low surface brightness haloes around
some Type I tails, as in Figure 2.  In addition, the inner radio lobes
 of M87, another high-RM central source, sit in a much larger extended
halo (Owen, Eilek \& Kassim 
2000).  For such sources, the interface between the radio plasma and
the ICM is not well-defined, thus shear-driven instabilities may not
develop.

\subsubsection{Magnetic fields throughout the cluster}

An alternative picture is that all clusters contain magnetized
plasma.  Conditions derived for cluster cores (such as the order scale
and fractional energy density of the field) should then be typical of
the entire ICM in the cluster.   

If a cluster-wide magnetic field exists, it may well have a uniform
strength within the gas core, but it is  very likely that the  field
strength falls off on larger scales.  For instance, if plasma dynamics
keep the 
pressure ratio $\beta = p_B / p_g$  constant throughout the cluster,
we would have $B \propto (nT)^{1/2}$.  It follows that the RM of
cluster-center sources will be dominated by conditions in the gas
core; outer regions  of the cluster will contribute little.   

We know of nothing in the data which rules this out.  We will show
below that the pressure ratio is typically $\beta  \sim O(.1) - 
O(1)$ in the magnetized filaments which contribute to the observed RM.
This seems inconsistent with both radio halo and statistical RM
studies, which suggest $\mu$G fields, and $\beta \ll 1$ on
cluster-wide scales.  However, both these observations are also
consistent with higher large-scale $B$ and $\beta$ values, if the
high-field regions do not fully fill the ICM.

\subsubsection{Cluster-wide fields amplified by cluster-center effects}

There is a third possibility.  It may be that the entire cluster
plasma is magnetized, but at a lower level, so that $\beta \ll 1$.
This weak, cluster-wide field could be 
enhanced to $\beta \sim 1$ levels in the region of the central radio
source. There are two ways in which this might happen.

One way to amplify the ambient field is compression in a
cooling-driven inflow.  Nine of the 12 high-RM central  sources  sit
in the centers of strong cooling cores.  If the short central cooling time in
these cores does lead to an inflow or collapse of gas toward the
center, any flux-frozen cluster field will be amplified by the density
increase (\eg, Soker \& Sarazin 1990, or Garasi 2001).
The ambient field can also be amplified by the interaction of 
the central radio galaxy with the local ICM. We expect a central radio
source to deposit a 
significant amount of energy into the local ICM.  This will drive
turbulence which can act as a local dynamo, amplifying the
cluster-wide field to higher levels.

We know of nothing in the data which rules out this picture.  It is in
reasonable agreement with the radio halo and statistical RM data. 
Thus, we conclude 
that we cannot discriminate between a strong, cluster-wide field, or
an enhanced cluster-center field, as the origin for the observed
central Faraday rotation.     

\subsection{Possible Magnetic Field Geometry}

The ICM magnetic field is unlikely to  be uniform and unidirectional
throughout the cluster.  Some authors propose large-scale order;
others propose small-scale randomness.  

We are aware of two models which predict large-scale order in the ICM
field.  Soker \& Sarazin (1990) proposed that a radial cooling flow
would stretch out initially random, turbulent fields into
approximately radial ``field loops''.  This structure would give a
characteristic RM signature: cells of alternating RM sign (assuming
the cells are resolved, so that beam depolarization is not an issue),
with amplitude decreasing with projected distance away from core (as
the radial field rotates away from the line of sight). We do not see
evidence of this in our data.  Eilek (1991) noted  that a
turbulent dynamo with strong helicity could maintain a field with
large-scale order.   The simplest solution for clusters is probably an
axisymmetric field in spherical geometry.   Such a field can be
described as sets of toroidal flux tubes.  The  RM signature would be
a few large-scale patches of ordered RM, both positive and negative.
This is consistent with, but not required by, our data. 

Alternatively, an ICM field supported by turbulence may have a disordered
structure. An older approach to turbulent fields assumes many
space-filling   `cells'' (\eg, Burn 1966).  Such a field would result
in a RM image which also appears random, with alternating ``cells'' of
positive and negative RM.  This picture does not seem consistent with
our images.  In addition, newer work on turbulent magnetic fields
suggests a different picture.  Numerical simulations reveal that  MHD
turbulence is often characterized by 
intermittency in the magnetic field.  That is, elongated, high-field
flux ropes are separated by low-field regions.  This  
geometry was discussed analytically by Ruzmaikin, Sokoloff \& Shukurov
(1989), and has also been seen in two- and three-dimensional MHD
simulations (\eg, Menguzzi, Frisch \& Pouquet 1981;  Biskamp 1993, \S
7.7; Kinney, McWilliams \& Tajima 1996; Miller \etal\  1996).  The
plasma density inside these flux ropes may be lower than that in the
surroundings, to maintain a pressure balance.  The
observational signature of an intermittent field would be high-RM
patches, possibly separated by lower-RM regions.  

Looking to the data, we note that all of the sources show evidence of
ordered RM patches, which may or may not alternate in sign, on scales
$\sim 1-10$ kpc.  This seems consistent with the intermittent field
picture.   In addition the larger  sources (A400 and A2634, also Hyd
A, Taylor \& Perley 1993) show evidence of larger-scale structure:
the apparent relation of the RM in A400 and A2634 to the 100-kpc scale
structure of the X-ray gas core, and the side-to-side asymmetry, on a
50-kpc scale, of Hyd A.  The data thus suggest a larger order scale
may exist which is not sampled by the smaller sources.   

To analyze our data, we therefore choose a model in which 
high-field flux ropes are surrounded by a lower-field background
plasma.  We do not have a good model for the larger order scale, but
it is not critical for our analysis here.  Our images of A400 and
A2634 show patches with scales $\sim  
10$ kpc, which we identify as the transverse size of the flux ropes. 
We note that such a geometry can have a large (area) covering factor
while having a small (volume) filling factor.  To show this, let the 
filaments have radius $r$ and length $L$, and immerse $N$ filaments 
in a region of radius $R$ (the gas core).  The covering factor is the
ratio of projected filament area to total projected area: 
$cf \simeq N rL / R^2$.  The
fractional volume occupied by the filaments is the filling factor, $ff
\simeq N r^2 L / R^3$.  The lack of low-RM holes in the images
requires that $cf \gtw 1$;  this is consistent with $ff \ltw 1$  when
$r < R$.    

\section{Results:  Magnetic Field in Cluster Cores}

In this section, we first analyze the field in our two clusters.  We
then put our results in the context of all nearby cluster-center
sources which have been studied in RM. 

\subsection{The Magnetic Field in A2634 and A400}

We refer to Table 1 to estimate the characteristic density 
and  temperature in the inner $\ltw 100$ kpc of each cluster.  For
Abell 400, we take $n \simeq 0.0021$ cm$^{-3}$, and $T \simeq 1.5$ 
keV; for Abell 2634 we take $n \simeq 0.0016$ cm$^{-3}$, and $T \simeq
1.5$ keV.  Referring to figures 7 and 8, we take 50 \radm\ as a 
typical RM for A400, and 65 \radm\ as typical for A2634.

\subsubsection{The Typical Magnetic Field}

We explore
three simple pictures: a thin skin on the radio source, a single
magnetic feature somewhere in the region of the source, and a fully
magnetized cluster core.   For each picture, we use 
equation (2) to derive $\Bparmean$, using the typical X-ray gas
densities given above. We estimate the magnetic pressure as $p_B
\simeq 3 \Bparmean^2 / 8 \pi$, and compare it to the gas pressure,
$p_g = n k_B T$.  Our results for each model are
summarized in Table 2.

We start by testing the model in which the  RM arises from a skin of
the radio source, in 
which magnetic field from the source has mixed with ambient ICM.  We
take the skin thickness to be 10\% of a typical radius of the radio
tail, and we estimate the latter very simply from the inspection of the
radio images.  For  3C75
we take $L_{RM} \sim 1.2$ kpc, and for 3C465 we take $L_{RM} \sim
0.7$ kpc. From the results in Table 2, we see that this picture
requires  quite high fields, of order tens of $\mu$G.  Such strong
fields would be be at a much higher pressure than the ambient gas,
indicating that such a shell is unlikely to be long-lived. 

It seems more likely that the magnetic field is associated with the
X-ray loud gas in the cluster core.  In this case, we assume
magnetic field is intermittent, with filament radius equal to the 
projected scale on which the RM is ordered. We thus take  $L_{RM} \sim 10$
kpc for A400, and $L_{RM} \sim 20$ kpc for A2634. 
We cannot predict the filling factor for the magnetic features.  We
therefore use the  two extremes of a single filament along the line of
sight, and a fully filled core.  A single filament requires a field
$\sim 2-3 \mu$G in each case, and predicts a magnetic pressure which
is interestingly close ($\sim 60$\%) to the thermal pressure in each
case.  
The other extreme assumes a filled core.  Comparing the gas
core radii (300 kpc for A400, 400 kpc for A2634) to the filament
diameters, and assuming the interfilament distance is  a few times the
filament diameter, we estimate typically $O(10)$ filaments would lie
along the line of sight through each  core.  The most probable
observed RM will be the RM of a 
typical magnetic feature, increased by the square root of the number of
such features superimposed along the line of sight. In this case, the
magnetic field is reduced by a factor of $\sqrt{10}$, and the
pressure ratio by a factor of 10, compared to the single-filament
case.   This limit reduces the magnetic
pressure to a few percent of the thermal pressure.  Both limits are
shown in Table 2.

\subsubsection{Extrema and Gradients in the RM}

We used the arguments above to estimate ``typical'' RM values and
magnetic field  strength.   Two further points are worth noting. 

First, the RM is not uniform in either object; local, high-RM patches
can be found (Figures 9 and 10). These patches are not generally
associated with any significant feature in the radio-loud plasma. 
These patches have larger RM, and smaller size, than is typical of the
rest of each source.  To analyze this, 
we assume these patches are unique features 
projected on the source, with line-of-sight scale equal to transverse
scale. We take $L_{RM} \sim 10$ asec $\sim 4.4$ kpc for A400 and $L_{RM}
\sim 7$ asec $\sim 4.1$ kpc for A2634.  This gives the final numbers
in Table 2; we 
see that high fields are required here. Unless these extrema have an
unusual geometry (with line of sight depth  greater than their
transverse scale) we find that they must be strongly overpressured
regions.   It follows either that they are magnetically self-confined,
or that they are short-lived  compared to the age of the radio galaxy. 

Second, the mean magnitude of the RM does not decay with radius as
rapidly as the density of the X-ray gas.  We can see this by comparing
Figures 7 and 8 to Table 1. 
For A400, the RM stays at a constant magnitude, while the density
drops by a factor of $\sim 4$ over the inner 100 kpc, and the
projected column density  drops by a factor $\sim 2$ over the same
range.  For A2634, the RM drops by a factor $\ltw 3$ in the inner 100
kpc, while the density drops by $\sim 8$ and the column density by a
factor $\sim 5$.  Thus, in each cluster the value of $\langle
B_{\parallel} \rangle L$ must increase, by a factor of a few, over the
inner $\sim 100$ kpc. 

\subsection{Comparison to Other Cluster-Center Sources}

We thus conclude that magnetic fields play a significant role in the
cores of A400 and A2634.  Most imaging of cluster-center RM up to now
has addressed smaller linear sizes and strong cooling cores.
Comparing our two clusters to the full set studied so far shows that
significant magnetic fields are common in cluster cores, whether or
not cooling is important.
To demonstrate this,  we have combined our data with all
published RM images of nearby cluster center sources. This gives us a
set of 12 nearby clusters (we exclude high-redshift objects due to
poorer knowledge of the gas distribution).   Our compliation is
similar to that of Taylor \etal\ (1999), but we consider
only  cluster-center sources, and carry out a different analysis.

\subsubsection{Magnetic Field Analysis}

Given the variety of resolution of the RM images and, 
especially, of the X-ray images of the sources, we have only attempted
a ``low resolution'' treatment.  We repeat the analysis of \S 4.1.1,
using X-ray data  to estimate the gas density and temperature which
are ``typical'' of the volume close to the radio source. 
We assume that one filament, with transverse scale $\sim L_{RM}$, lies
along the line of sight. The smaller cores typical of these clusters
make it unlikely that more than a few such flux ropes lie along the
line of sight.   We put our detailed assumptions, and specifics for
each cluster, in the Appendix.    
In Table 3 we list  for each source:  the linear size of the radio
source over which RM has been measured, $D$;  the strength of the 
cooling core, estimated by $\dot M$, the mass inflow rate in 
a traditional cooling-flow analysis;  the 
typical RM value;  the scale length over which the RM is ordered;  the
typical density and temperature for the X-ray gas; our results for
$\Bparmean$; and our derived ratio of magnetic pressure to gas
pressure. 

Table 3 shows that there is a clear trend for the magnetic field to be
significant in these clusters.  All but three of the sources have $p_B /
p_g \gtw 0.1$, which we consider a reasonable threshold for the
magnetic field to have some importance in the gas dynamics.  Three of
the sources even have $p_B \ge p_g$. Thus, we conclude the magnetic
field is generally important in these cluster cores, and that $
p_B / p_g \sim O(0.1) - O(1)$ is a ``typical'' value of the
pressure ratio in the high-field filaments.  

Table 3 also shows that the magnetic field
correlates with the gas density in the core. 
This is a stronger result than a correlation of the RM
with the density.   The correlation of $B$ with density shows that the
magnetic field is closely connected to the dynamics of tye ICM.  We
return to this below, when we consider how the field is maintained.

\subsubsection{Caveats}

It is worth pointing out the uncertainties in our ``low resolution''
estimates of the cluster-center magnetic fields.

First, the X-ray deprojections contain uncertainties. The 
deprojected densities should be fairly robust, on and above the beam
size.   X-ray emissivities are not very sensitive to temperature, for
gas above $\sim 10^7$K,  and
the clusters in question do not deviate significantly from approximate
sphericity.  On scales less than the beam size, where we had to
estimate the density enhancement around a small radio source and
ignore central asymmetries, our numbers are of course less reliable.
In 
addition, the local temperature is uncertain.  This is true because
because older published work only gives a 
mean temperature for the cluster, and because cooler gas in the
cluster core can be hidden in the final X-ray image. Thus, the central
temperature, and thus gas pressure, may be significantly less than our
estimates here.  This would increase the plasma $\beta = p_B / p_g$.   

Second, we assumed the simplest geometries in deriving magnetic field
values.  We assumed a unity covering factor for magnetic features.
This can lead to an overestimate of the field strength. In addition,
we assumed the X-ray gas is well mixed in 
the high-field filaments.  This may not be the case, as magnetized
filaments may expel plasma in order to stay in pressure balance with
their surroundings.  This effect will give larger field values than we
have estimated.  Finally, we have used a ``typical'' RM value for each
source;  this ignores the smaller, high-RM patches which exist in
several of the sources.  As with A400 and A2634, high-field patches
may also exist in these clusters.

\section{What Maintains the Magnetic Field?}

We have shown that strong magnetic fields, $\beta \ltw 1$, are common
in the centers of rich clusters.  These fields require a driver.  
The small order scale, $\lambda_t \sim 10$ kpc, means that
reconnection will dissipate the field on a time $\tau_A \sim 10
\lambda_t / v_A$, if $v_A = B / (4 \pi \rho)^{1/2}$ is the Alfven speed
(\cf\ Parker, 1973, or Garasi, 2001). $\tau_A \sim 100$ Myr 
for typical cluster conditions.  Thus, magnetic driving is needed;
turbulence in the ICM is the most likely mechanism.  We also note the
correlation 
of the magnetic field and gas density (in Table 3) strongly suggests a
dynamical connection between the field and the thermal plasma.
We need to understand what drives the turbulence.  

\subsection{Turbulent support of magnetic fields}

We summarize the important parameters which allow quantitative
estimates of turbulent driving.  Let $v_t$ be the characteristic
turbulent velocity, and $\lambda_t$ be the characteristic scale,
probably the scale on which the turbulence is driven.    $E_t \simeq
\rho v_t^2 V$ is the turbulent energy 
in volume $V$.  Let $P_t$ be the driving power for that volume.
Viscous effects will dissipate the turbulence on a time $\tau_t \sim
{\rm few} \times \lambda_t / v_t$.  For times $t < \tau_t$, the
turbulent energy $E_t \sim \int P_t dt \sim P_t t$;  for times $t >
\tau_t$, the system reaches a balance at $E_t \sim P_t \tau_t$.  For
transonic turbulence, on small scales (say, $\lambda_t \sim 10$ kpc),
the decay time $\tau_t \sim 50 - 100$ Myr;  thus small-scale
turbulence must be continually replenished.  Alternatively, if
$\lambda_t \sim $ Mpc, as in merger driving (\eg, Norman \& Bryan
1999, or Roettiger \etal 1997)  we would not expect energy balance to
be established.     

If the turbulent fluid is a magnetized plasma, the mean magnetic field
will be amplified to approximate dynamic balance, 
\be
B^2 \ltw 4 \pi \rho v_t^2
\ee
This process needs only a very small seed field.  
Early-epoch galaxy activity is a possible source for seed
fields (\eg, Kronberg \etal 1999, V\"olk \& Atoyan 1999),
as is wind ejection from normal galaxies at the present epoch.   
Note that the high fields we find in cluster 	
cores, with $B^2 \sim 8 \pi p_g$, require that the turbulence be
transonic.    
Simulations (De Young 1980, 1992) show that the rate at which dynamo
amplification happens depends on the nature of the turbulence.  The
helicity of the 
turbulence is critical here.  (Helicity is defined as
$\mathbf {v} \cdot \nabla \times \mathbf{ v}$; a tornado, with flow
along its rotation axis, 
is a dramatic example of helical flow).  Simulations show that simple,
isotropic (non-helical) turbulence takes many characteristic times (30
- 100) to 
establish this balance.  Strongly helical turbulence is faster,
establishing dynamical balance in $\sim \tau_t$.  
The spatial structure of the field also depends on the structure of the
turbulence.  Isotropic, non-helical turbulence amplifies the
magnetic field on scales no larger than the driving scale.  Such
turbulence is probably intermittent, containing thin, high-field flux
ropes. Helical turbulence leads to strong fields which are ordered on
scales $\gtw \lambda_t$ (although it can take much longer  
than $\tau_t$ for significant power to cascade to these long
scales).  We cannot easily predict the amount of helicity in cluster
turbulence, although we expect some due to galaxy rotation and to angular
momentum associated with the cluster formation process.

\subsection{Maintaining a cluster-wide field}

\subsubsection{Galaxy Driving}

The motions of galaxies through the ICM will heat the ICM, and will
drive turbulence in the ICM at some level.  That turbulence will
support amplify an initial seed field to approximate equiparation
levels, as in (4).  As this must be occuring in all clusters, it
provides the minimum turbulent and field levels to be expected.  These
levels turn out to be very weak. 

We estimate the energetics as follows.  
Say that $N_g$ bright
galaxies (close to $L_*$) occupy a cluster core.  They transfer  
energy to the ICM over their gravitational radius, $r_g \sim 5$ kpc
(for $M_g \sim 10^{12}M_{\sun}$, and $v_g \sim 10^3$ km/s).
The rate of  energy transferred to the ICM  will be  $\rho v_g^3 \pi
r_g^2$, times some efficiency $\epsilon$.  If 
we assume energy equipartition, $M_g v_g^2 \simeq$ constant,
we find a driving power that is a strong function of galaxy mass:
$P(M_g) \simeq \pi \epsilon \rho G^2 M_g^2 / v_g $.

We can include a distribution of galaxy masses.
We take  $N(M_g) \propto M_g^{-5/4}, M_g <
M_*$, if $M_* \sim 10^{12}M_{\sun}$ is the cutoff mass.   
The total number of galaxies is $N_g = \int N(M_g) d M_g$. Energy
equipartition gives  $v_g(M_g) \sim \sigma 
( M_* / M_g)^{1/2}$, where $\sigma$ is the cluster velocity dispersion.
The net driving power is then
\be
P_g = \int_0^{M_*} N(M_g) P(M_g) d M_g  \simeq { \pi  \over 9} 
\epsilon N_g { \rho \over \sigma} G^2 M_*^2
\ee

We estimate the steady-state turbulence level by balancing this
driving against dissipaton:
\be
P_g \simeq P_t \simeq { \rho v_t^3 \over \lambda_t}
\ee
This relation gives us an estimate of the turbulent energy density,
$\rho v_t^2$.   From this, using (4), we find the mean magnetic field
that can be maintained by galaxy driving:
\be
B \simeq 0.26 n_e^{1/2} \left( \epsilon N_g\right)^{1/3} \left({  M_{12}
\lambda_{10} \over \sigma_e} \right)^{1/3} \quad \mu{\rm G}  
\ee

To put in numbers, we consider a very rich cluster, to make the best
case for galaxy driving.  We pick $n_3 = M_{12} = \lambda_{10} =
\sigma_e = 1$, and $\epsilon N_g = 10$.  This last is meant to 
describe a  rich cluster with moderate efficiency.  Under these
conditions, we estimate that galaxy driving will maintain turbulent
velocities $v_t \sim 40$ km/s, and $B \sim 0.5 \mu$G.  
This result shows (in agreement with some previous authors:  Goldman
\& Raphaleli 1991,   De Young 1992), that only a very small mean field
can be maintained by galaxy-driven turbulence.

\subsubsection{Driving by cluster relaxation and mergers} 

An alternative to galaxy driving is that the magnetic field is
maintained by turbulence driven by ongoing mergers. 
There is substantial evidence that not all clusters are  quiescent and
dynamically relaxed.  Many are still evolving dynamically, and some
are undergoing dramatic major mergers at the present epoch.
In addition, simulations suggest 
that minor mergers are common, with  infalling matter still
accumulating into the potential wells of large clusters.  

This can be a much stronger driver than the motion of
individual galaxies.  Norman \& Bryan (1999) suggest $10^{62}$ erg for
the kinetic energy of infalling subclusters in their ``minor merger''
simulations.  This value is comparable to the internal energy of a rich
cluster, and much greater than the energy supplied by galaxy driving
(equation 5) over the life of the cluster.   A cluster with a recent
merger may still contain several high-velocity clumps.   Each such
clump will energize the ICM in the same way as a large galaxy will,
but at higher levels.  

Mergers have also been suggested as the answer to a problem posed by
Eilek \etal\  (1984), and  Owen \etal\ (1985).  These papers showed
that the the radio sources in A2634 and A400 cannot be bent by the
motion of their parent galaxies relative to a static ICM in a static
cluster potential.  The situation could be 
resolved if the ICM exhibits large-scale transonic flows,
such as can be attained in a favorable merger event (Roettiger \etal
1997).  It may be that these  clusters have 
undergone merger events fairly recently, and that turbulence driven by
the merger is responsible for the magnetic fields therein.  

Thus, it is plausible that ongoing mergers drive cluster-wide
turbulence strongly enough to maintain transonic turbulence, and
dynamically significant magnetic fields, throughout the cluster.  We
caution, however, that this hypothesis has not yet been established
quantitatively.   Mergers occur on large scales ($\sim$ Mpc).  We 
would expect the characteristic size of merger-driven
turbulence to be large as well. This seems
to conflict with the data which show $\sim 10$ kpc order scales in the
RM, suggesting 
significant turbulent power on these scales.  It has not yet been
established whether or not merger-driven turbulence can cascade
effectively to these scales.  (Numerical simulations are  not yet able
to evaulate 
turbulent and magnetic field levels on scales small compared to the
cluster or merger;  they are limited both by finite resolution and by
the difficulty of including dissipative and resistive effects).  
Another concern is that many of the 
clusters with high RM values and high $B$ fields, also contain strong
cooling cores.  There are problems with the suggestion that these
clusters have recently undergone significant mergers.  A strong merger
will heat the central ICM (as in Roettinger \etal\ 1997, or
Schindler \& Prieto 1996), and disturb the underlying gravitational
potential.  It is not obvious that the relaxed profiles and lower
central temepratures characteristic of cooling cores are compatible
with recent mergers in these clusters.

\subsection{Cluster Core Amplification}

\subsubsection{Cooling Flows as a Driver}

Another possibility is suggested by the fact that most of the sources
in Table 3 sit in the center of strong cooling cores.  It may be 
that the high magnetic fields in these clusters  are local to the
cluster core, and are due to a cooling flow. If such flows exist, they
will  
enhance an initially weak magnetic field by simple compression.  
Soker \& Sarazin (1990) found that an initial field at the $\mu$G
level would reach dynamic balance within the inner $\sim 10$ kpc of a
standard cooling-flow model.  Recent numerical simulations, such as
the one-dimensional, tangled-field models of  Garasi  (2001), find that 
fields on the order of tens of $ \mu$G are very possible in strong
cooling cores.  

Field enhancement by a cooling-driven inflow may indeed be
important in some of the high-RM clusters.  However, we do not believe
that current models of quasi-spherical cooling flows can explain all
the data. Three of the sources in Tables 3 and 4, including the two
new  new sources in this paper, are not in strong cooling  cores.
There may be a small X-ray excess in the very inner region of both
A400 and A2634;  but there  is no evidence for a cooling inflow on the
$\sim 100$ kpc scale where RM is observed, and significant fields are
inferred, in these two clusters.

Furthermore, it is becoming less clear that simple, slow inflows
described in current models really characterize clusters with cooling
cores.   As we describe in the next section, evidence is accumulating
that  active central  galaxies deposit a significant amount of energy
into the cluster core.  Such energy input will heat and disturb the
gas, providing both kinematic and magnetic pressure support, and
possibly leading to a cycle of infall and outflow (regulated by the
lifetime of the radio source), on times short compared to the
life of the cluster (\eg, Binney 1999).   Thus, a cooling-driven
collapse may contribute to the magnetic fields seen in many of the
clusters studied do far, but we do not think it is the full solution.

\subsubsection{The radio source as a driver}

High magnetic fields can also be local to the
cluster core is if they are maintained by the radio galaxies
themselves.  Recent X-ray work provides clear evidence of such
interaactions.  Examples are 3C388 in A2199 (Owen \& Eilek 1998), M87
(Owen \etal\ 2000;  Harris \etal 2000), or Hydra A (McNamara \etal, 2000). 
Energy arguments, presented below, show the radio source can have
significant impact on the local ICM.  The connection
will be hydrodynamic:  the radio-jet plasma will deposit energy into the
local ICM, both as heat and as bulk kinetic energy (``turbulence'').
Such radio-source-driven turbulence can act as a local dynamo,
amplifying the cluster-wide field to larger levels.  This may be the
origin of the $\beta \sim 1$ fields found near cluster-center
sources. 

In order to test this idea, we carried out simple sums for the set of
ten sources in Table 3.  We compiled 
radio data from the literature:  the largest size, $D$, over which RM has
been measured;  the bolometric radio power 
$P_{rad}$ (assuming a 10 MHz - 10 GHz  band);
and the minimum pressure internal
magnetic field, $B_{mp}$ (although the diversity of observations and
resolution makes 
this quantity only a very approximate measure of the internal energy
density of the sources).  These quantities are listed in Table 4;
details and references  are in the Appendix).  Table 4 shows that 
$B_{mp}$ and $\Bparmean$ are 
somewhat correlated.  This suggests an approximate dynamic balance
between the pressures inside the radio source and in the local ICM. 

One could also explore this picture by considering timescales. 
Let the radio beam carry power $P_b$, and assume most of this is 
deposited in the local ICM.  
It is simple to determine how long it
takes for the beam to deposit an 
amount of energy equal to the 
total thermal energy in the central volume {\it for which the RM has
been measured} (this last is a minimum energy requirement).
This is, 
\be
\tau_E = ( 8 \pi / 3) n k_B T D^3 / P_b
\ee
We assume the radio jet carries total beam power  $P_b = 100 P_{rad}$, 
The final column of Table 4 shows that this times can be short.  The
smaller sources, with their smaller associated volumes, have
$\tau_E \sim $ few Myr.  This is quite a short time; one would
clearly expect these sources to affect their immediate surroundings.
The two largest ones have $\tau_E \sim 200-300$ Myr.  This is a longer
time, but not significantly longer than the ages estimated for large, tailed
radio galaxies. Thus, it may be that the high magnetic
fields we observe are only local to the cluster core, and are
maintained by turbulence which is driven by the radio galaxy itself.

\section{Summary and Conclusions}

In this paper we have conbined new and existing Faraday rotation data to 
study the magnetic field in the cores of clusters of galaxies.  

We presented new observations of two large, tailed radio galaxies
at the center of non-cooling clusters.  We used these data together
with X-ray 
data to estimate the field strength,  and found dynamically interesting
fields exist in the inner $\sim 100$ kpc of A400 and A2634. 
We also used data from the literature to carry out a simple RM
analysis for the 12 clusters with RM data for cluster-center sources.
We found that significant magnetic fields (with $\beta \sim 0.1 - 1$)
are common in cluster cores.  In our analysis we assumed the magnetic
field has a filamentary structure, and that the RM is dominated by the
high-density (and probable high-field) gas core of the cluster.   

We then considered possible origins and drivers for the field.  Galaxy
motion is only a weak driver.  The turbulence that galaxies create in
the ICM may maintain fields at a sub-microgauss level, but cannot
account for the stronger fields detected in rotation measure.  We do find
three possible mechanisms to support the larger fields.

1.  Ongoing mergers, at least minor ones, are likely in many clusters.
Such mergers should be able to drive strong turbulence in the ICM, and
thus maintain interesting levels of magnetic field.  The asymmetric inner
gas distribution in A400 and A2634 may be due to recent merger activity
in these two clusters. We caution, however,
that this model cannot easily be quantified to predict specific field
levels;  and note that it may not explain the fields found in strong
cooling cores.

2.  Cooling-driven collapse of the ICM can amplify a weaker cluster-wide
field in the central cooling core.  As previous authors have noted, this 
may 
account for the high fields found in the cooling cores previously studied. 
It clearly is not relevant, however, to the two clusters presented in
this paper, which do not sit in cooling cores.

3.  The central radio galaxy may also play a role in maintaining a strong
local field.  Some part of the jet power will be deposited
in the nearby ICM, driving turbulence as well as heating that ICM.  Simple
energetic considerations show that strong turbulence, and field, levels
are possible local to the radio source.

We cannot discriminate between these three models.  We suspect that all are
operating at some level, in at least some clusters.  What we can conclude
is that the intracluster plasma is not as simple as many of us once
thought.

\begin{acknowledgements}
We are pleased to acknowledge stimulating conversations with   Tracy
Clarke, Dave De Young, Chris Garasi, 
Tomislav Markovic, and Mark Walker, over the course of this
work. Comments from an anonymous referee improved the structure and
organization of the paper.   We
are also grateful to Paddy Leady for the initial stimulus to carry out
the observations, and to Qiang Wang for help with the X-ray data.
This work was partially supported by NSF grant AST-9720263. 
\end{acknowledgements}

\begin{appendix}

\section{X-ray Analysis and Individual Clusters} 

We have detailed
X-ray deprojections for the two sources in this paper, for A2199 from
Owen \& Eilek (1998), and for M87
(Owen, Eilek \& Kassim 2000).  The extent of the radio sources sampled
by Faraday rotation  in  A400, A2199 and A2634 is larger than the X-ray
beam.  For these sources we used the deprojections to estimate ``typical'' 
densities across the RM region.  The extent sampled in RM in M87 and
in the 3C129 cluster are  comparable to the X-ray beam, so we used the
central density from deprojections in the literature.  For the other 7
sources, we have only  published X-ray analyses. These give  
X-ray temperature, and also either the  gas density in the central
X-ray bin, or the cooling time in that bin (which, with an assumed
temperature, can be converted to number density;  we use the analytic
cooling curve of Westbury \& Henriksen 1992 
to do this).  All of these sources are smaller than the X-ray
resolution element, and all are in strong cooling cores.  The local
gas density at the radio source will thus be larger than the mean over
the Xray beam.  To estimate this correction, we used the
cooling-cluster density models of Eilek \& Markovic(2001), and the
estimated size scale of the X-ray gas for each particular
cluster, as described in that paper. Our estimated
corrections were factors from 2 to 4 in all cases.  Finally we
converted all estimates to our assumed ${\rm H}_o = 75$ km/s-Mpc.  

In the following we give relevant details for each cluster. 
We use the following abbreviations:  KPTW, Kellerman, Pauliny-Toth \&
Williams (1969); ES91, Edge \& Stewart (1991); ESF92, Edge, Stewart \&
Fabian 1992; W97, White \etal (1997); P98, Peres \etal (1998).  

\bigskip
{\it Abell 400} contains 3C75, a Wide-Angle Tail radio
source. KPTW 
give fluxes and spectral indices; the source can be fitted by a single
power law, $\alpha = - d \ln S_{\nu} / d \ln \nu \simeq 0.8$ from 10
MHz to 10 GHz. Owen \etal\
(1985) present the radio image; the Faraday data and cluster magnetic
fields are discussed in the present paper.  Beers \etal\ (1992) present
EINSTEIN data;  the ROSAT data is presented in this paper.  In \S 4 we
take the typical $T \sim 1.5$ keV.  This cluster has a broad gas
core, making it a non-cooling cluster, but also has  a weak central
X-ray excess which may be a small cooling core.

\bigskip
{\it Abell 1795} contains the tailed radio source 4C46.42. van
Breugel, Heckman \& Miley (1984) present radio fluxes and estimate
$\alpha \simeq 1.0$ over the radio range.  Ge \& Owen
(1993) present the radio image and Faraday rotation data.
We use X-ray data from P98, W97, ES91;  and estimate the density
close to the radio source $\sim 4$ times the mean density in the inner
(IPC, PSPC) X-ray beam.

\bigskip
{\it Abell 2052} contains the diffuse radio source 3C317.  Zhou \etal
(1993) compile fluxes, and KPTW give $\alpha \simeq 0.92$ at low
frequencies; the source steepens above $\sim 300$ MHz.   Ge \& Owen 
(1994) present a radio image, as do Zhou \etal.
We use X-ray deprojections from W97, P98, and estimate the density
close to the radio source $\sim 2.4$ times the mean density in the inner
(IPC, PSPC) X-ray beam.

\bigskip
{\it Abell 2029} contains a tailed radio source, PKS 1508+059.
Arnaud (1985), also Jones \& Forman (1984) use EINSTEIN data to
estimate gas densities and temperature. Recent radio data
are presented by Taylor, Barton \& Ge (1994), who also discuss the
cluster magnetic field.  This is a steep spectrum source, $\alpha
\simeq 1.5$ down to 80 MHz (Taylor \etal 1994).  We use X-ray
deprojections from ES91, ESF92, and P98,  and estimate the density  
close to the radio source $\sim 2$ times the mean density in the inner
(IPC, PSPC) X-ray beam.

\bigskip
{\it Abell 2199} contains the diffuse radio source 3C338.  KPTW give
fluxes; this is a steep spectrum source, $\alpha \simeq 1.09$ at low
frequencies.  Ge \& Owen (1994) present the radio image.  Owen \& Eilek
(1998) present and deproject ROSAT data and use Faraday data to infer
the magnetic field.   We use their X-ray deprojection, and estimate $T
\sim 2 $ keV based on Thomas, Fabian \& Nulsen (1987).

\bigskip
{\it Abell 2634}  contains 3C465, a Wide-Angle Tail radio source.
The spectrum can be described by a single power law, $\alpha \simeq
0.8$ (KPTW).   Eilek
\etal\ (1984) present radio images and the EINSTEIN data; the Faraday
data and ROSAT data are discussed in the present paper.
In \S 4 we take the typical $T \sim 1.5$ keV.  This cluster has a broad gas
core, making it a non-cooling cluster, but also has a weak central
X-ray excess which may be a small cooling core.  

\bigskip
{\it Abell 3526} is also called the Centaurus cluster.  The 
small, tailed radio source 1246-410 is associated with the central
galaxy.  Taylor \etal (1999) present the radio image, which suggests
that the faint radio tails extend beyond the detected limits, and also
the RM.  We have not found other radio data;  we assume the total flux
is twice that given by Taylor \etal\, and $\alpha \simeq 1.0$ between
10 MHz and 10 GHz to estimate total power.  The Xray data is from P98
and Churazov \etal\ (1999).  

\bigskip
{\it Abell 4059} contains a tailed radio source PKS 2354-350.  Taylor
\etal (1994) present radio data; this is a steep spectrum 
source, $\alpha \simeq 1.43$ down to 30 MHz. Schwarz \etal\ (1991) present
EINSTEIN X-ray data. We use X-ray deprojections from ES91, ESF92, 
P98,  and estimate the density close to the radio source to be $\sim
2$ times the mean density in the inner (IPC, PSPC) X-ray beam.

\bigskip
{\it The 3C129 cluster} contains 3C129.1, a small tailed radio source
associated with the central cD galaxy, in addition to the famous
tailed source 3C129. We consider the former source.  Taylor \etal\
(2001) present  5 GHz radio data and RM data.  We use the NVSS to
estimate total flux at 1.4 GHz, assume $\alpha = 1.0$ between 10 MHz
and 10 GHz to estimate total radio power.  The X-ray data are from
W97, Taylor \etal; see also Leahy \& Yin (2000).  This cluster has a
broad gas core, making it a non-cooling cluster, but also has a weak
central X-ray excess which may be a small cooling core.   

\bigskip
{\it Cygnus A} is a classical double radio source in the center of an
anonymous cluster.  It can be described approximately by a power law
with $\alpha \simeq 0.66$ below $\sim 2$ GHz, and appears to turn over
around $\sim 25$ MHz (Parker 1968; Baars \&
Hartsuijker 1972). Dreher, Carilli \& Perley (1987) present the
Faraday rotation data, and discuss cluster magnetic fields (for radio
images, \cf\ Perley, Dreher \& Cowan 1984). We use X-ray deprojections
from ES91, W97, P98,  and estimate the 
density close to the radio source to be the mean density in the inner
(IPC, PSPC) X-ray beam. 

\bigskip
{\it Hydra A} is a tailed radio source in the center of an anonymous
cluster.  Taylor \etal\ (1990) present the radio data; Taylor \& Perley
(1993) 
discuss the cluster magnetic fields.  Fomalont (1971) gives a 1400 MHz
flux. The  spectrum is a steady power 
law, $\alpha \sim 0.9$, over the radio range (Taylor
2000, private communication). We use X-ray deprojections from ES91, ESF92, 
P98,  and estimate the density close to the radio source to be the
mean density in the inner (IPC, PSPC) X-ray beam.   

\bigskip
{\it Virgo A} is the central radio source (M87) in the Virgo cluster.
Its spectrum is a good power law, $\alpha \sim 0.8$ throughout the
radio range (\eg, KPTW). To be consistent with our application here,
we do not include flux above 10 GHz; thus our bolometric power is
smaller than that quoted by Turland (1975), who extrapolates to
$10^{15}$ Hz. Recent radio images of the inner halo are in Hines, Owen
\& Eilek (1989); Owen Eilek \& Keel   (1990) presented the Faraday
image of the inner halo and used EINSTEIN data to infer the 
magnetic field.  N\"ulsen \& Bohringer (1995) present and deproject
ROSAT PSPC data; Owen \etal (2000)  convert their results to $H_o =
75$km/s-Mpc, which we use here. 

\end{appendix}


\clearpage \plotone{f1.eps}

\figcaption[fig1]{Abell 400 and 3C75.  The contours show the X-ray
emission from A400, observed with the ROSAT PSPC, convolved to 80 asec
$\sim 35$ kpc resolution.  The grey scale shows a 327 MHz VLA image of
3C75.  The image is 900 kpc on a side.}   

\clearpage \plotone{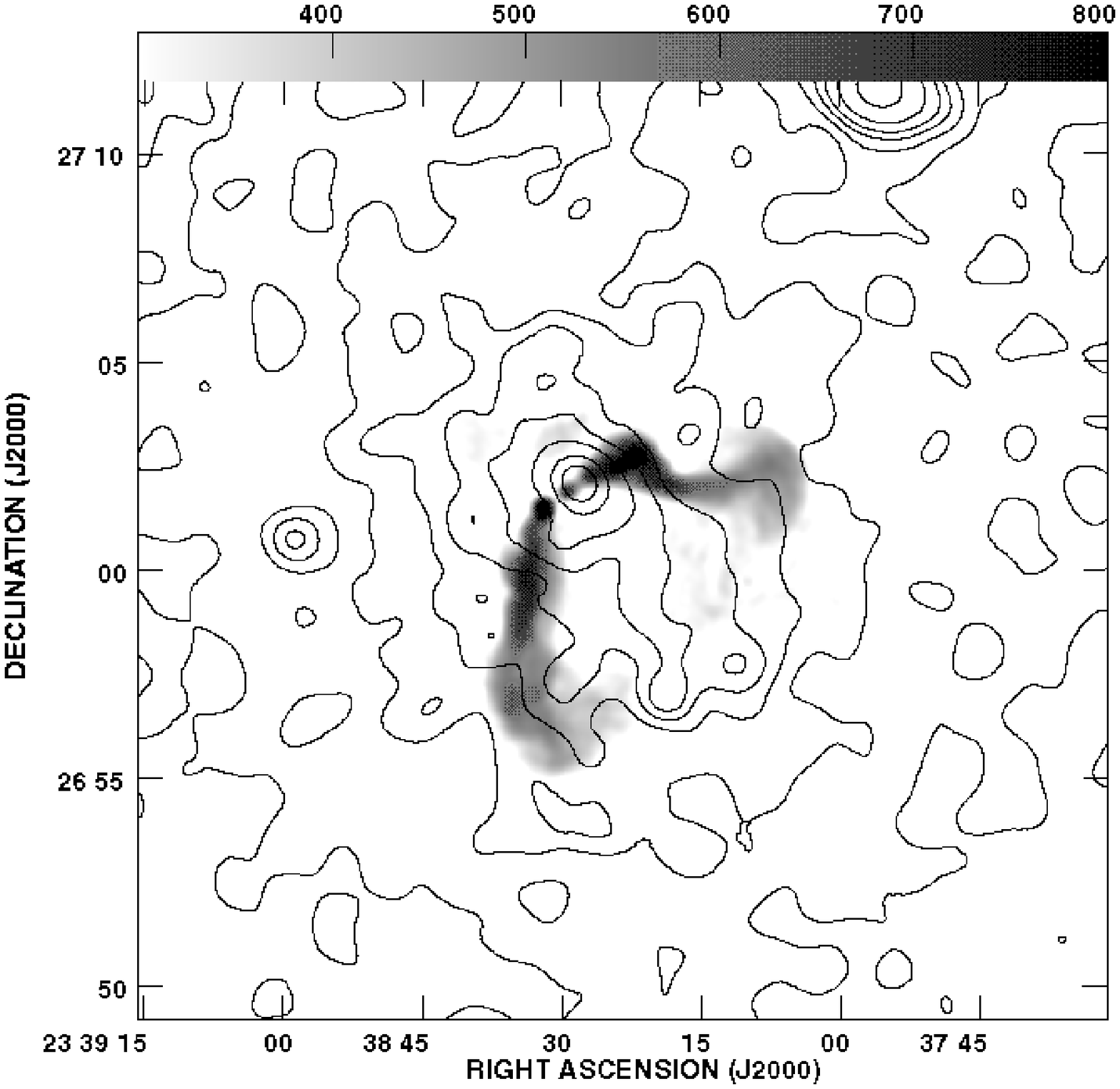}

\figcaption[fig1] {Abell 2634 and 3C465.  The contours show the X-ray
emission from A2634, observed with the ROSAT PSPC, convolved to 60
asec $\sim 35$ kpc resolution.  The grey scale shows a 327 MHz VLA
image of 3C465.  The image is 900 kpc on a side.   }

\clearpage \plotone{f3.eps}
\vspace{-0.8in}
\figcaption[fig3] {Magnetic field vectors, projected on the sky, for
3C75.  Faraday rotation has been accounted for, so that the intrinsic
field direction can be determined. All vectors have the same length;
no attempt has been made to display the fractional polarization.   The
contours refer to the 5 GHz radio image. } 

\clearpage \plotone{f4.eps}
\vspace{-0.8in}
\figcaption[fig4]{ Magnetic field vectors, projected on the sky, for
3C465.  Faraday rotation has been accounted for, so that the intrinsic
field direction can be determined. All vectors have the same length;
no attempt has been made to display the fractional polarization.   The
contours refer to the 5 GHz radio image. } 

\clearpage \plotone{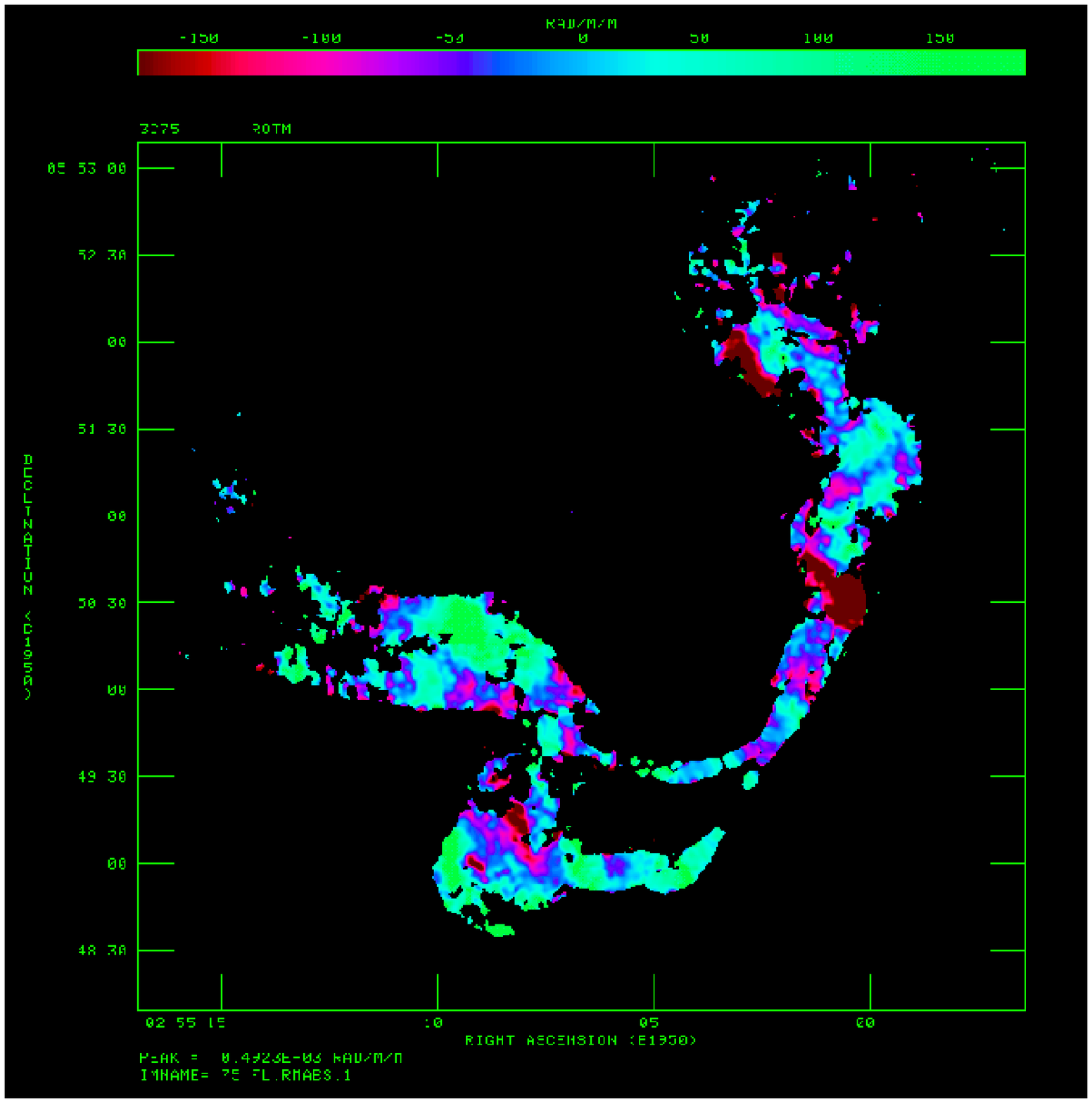}

\figcaption[fig5] {Rotation measure distribution in Abell 400.  The
colors are chosen to distinguish positive (green and light blue) and
negative (red and dark blue) values.  One arcmin is 26.5 kpc.}

\clearpage \plotone{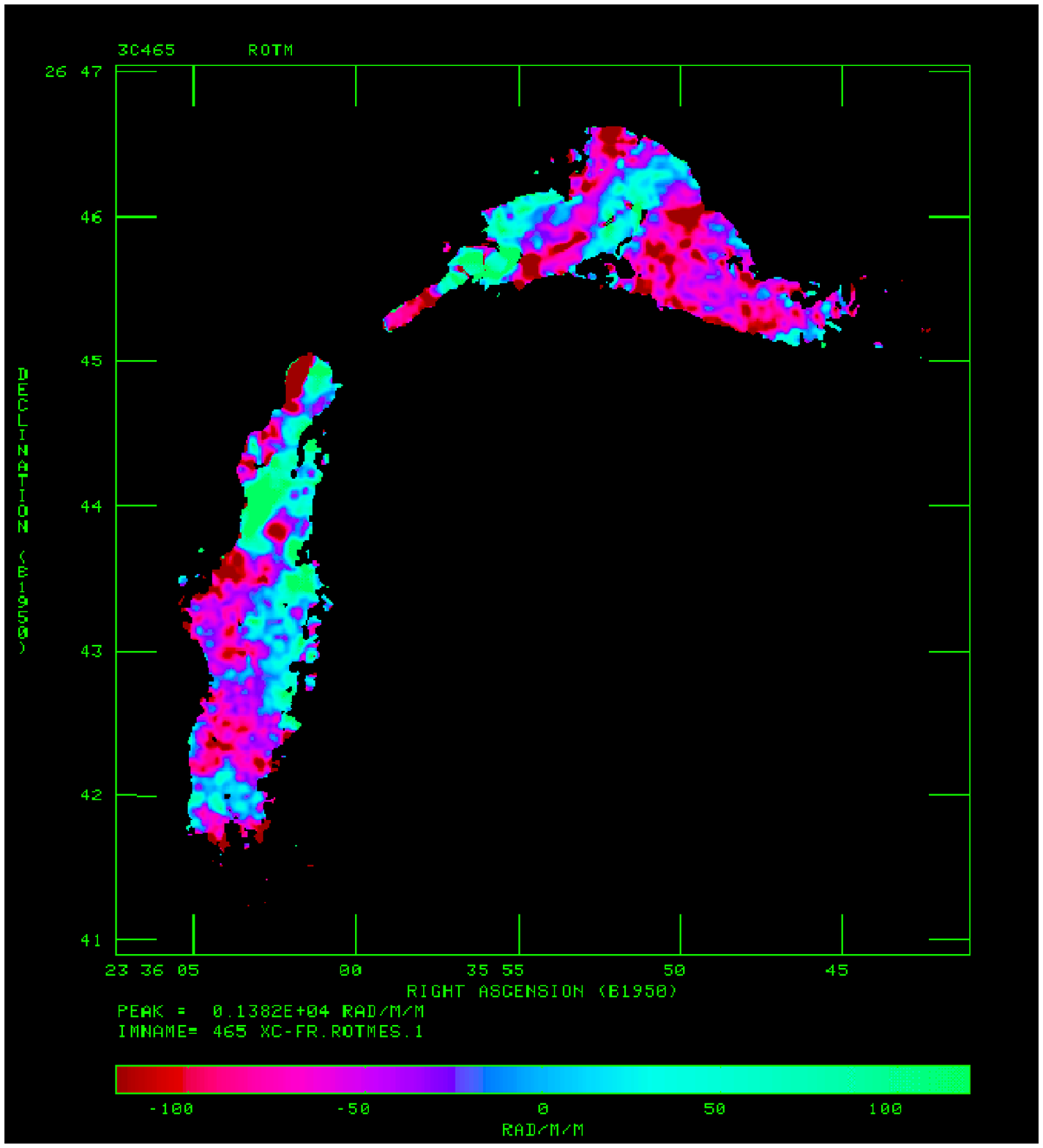}

\figcaption[fig6]  {Rotation measure distribution in Abell 2634.  The
colors are chosen to distinguish positive (green and light blue) and
negative (red and light blue) values.  One arcmin is 35.2 kpc.}

\clearpage \plotone{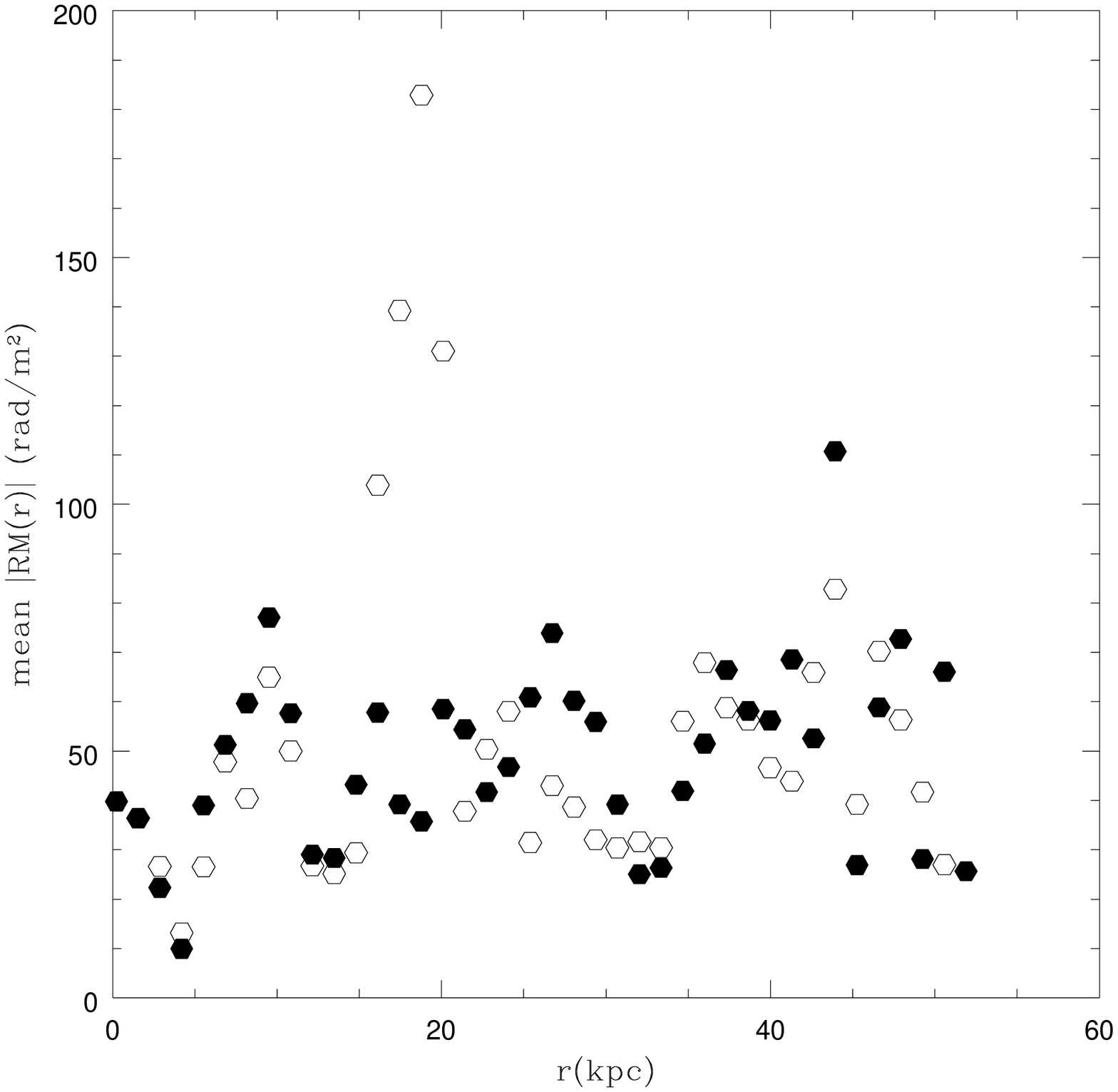}

\figcaption[fig7] {Absolute value of the rotation measure in Abell 400,
as a function of distance.  The points are means within circular rings
centered on the radio core.  The east tail is shown as solid points,
and the west tail as open circles.  The strongly negative RM patch on
the west tail is apparent at $\sim 150$ \radm.}

\clearpage \plotone{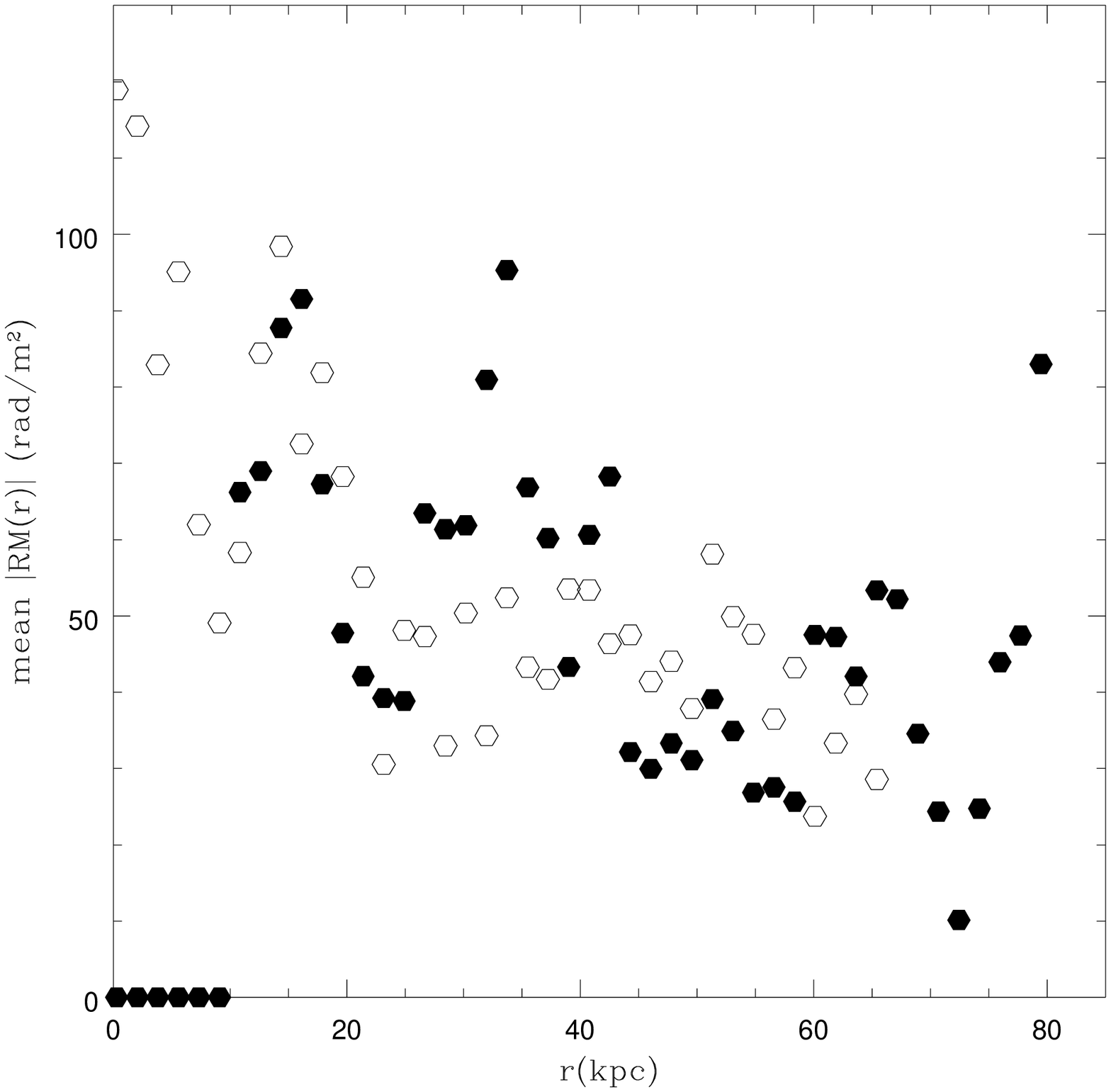}

\figcaption[fig8] {Absolute value of the rotation measure in Abell 2634,
as a function of distance.  The points are means within circular rings
centered on the radio core.  The east tail is shown as solid points,
and the west tail as open circles.  Note the decay of $|{\rm RM}|$
with distance in this cluster.}

\clearpage \plotone{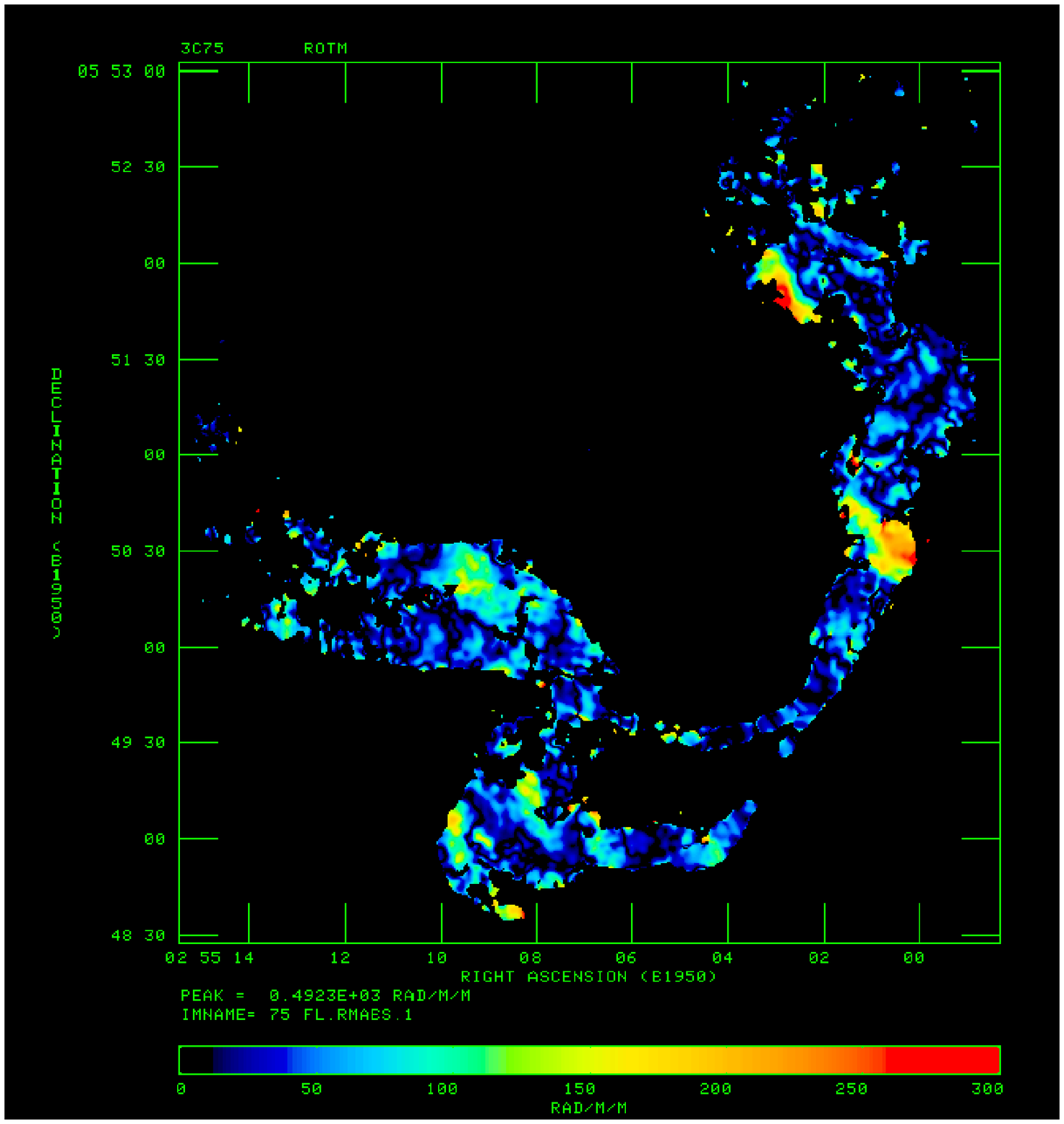}

\figcaption[fig9] {Absolute value of the RM in 3C75.  The colors are
chosen to highlight extreme values of the RM:  the bright yellow patches on
the west tail reach -300 \radm\ (northerly patch) and -200 \radm\
(southerly patch). }

\clearpage \plotone{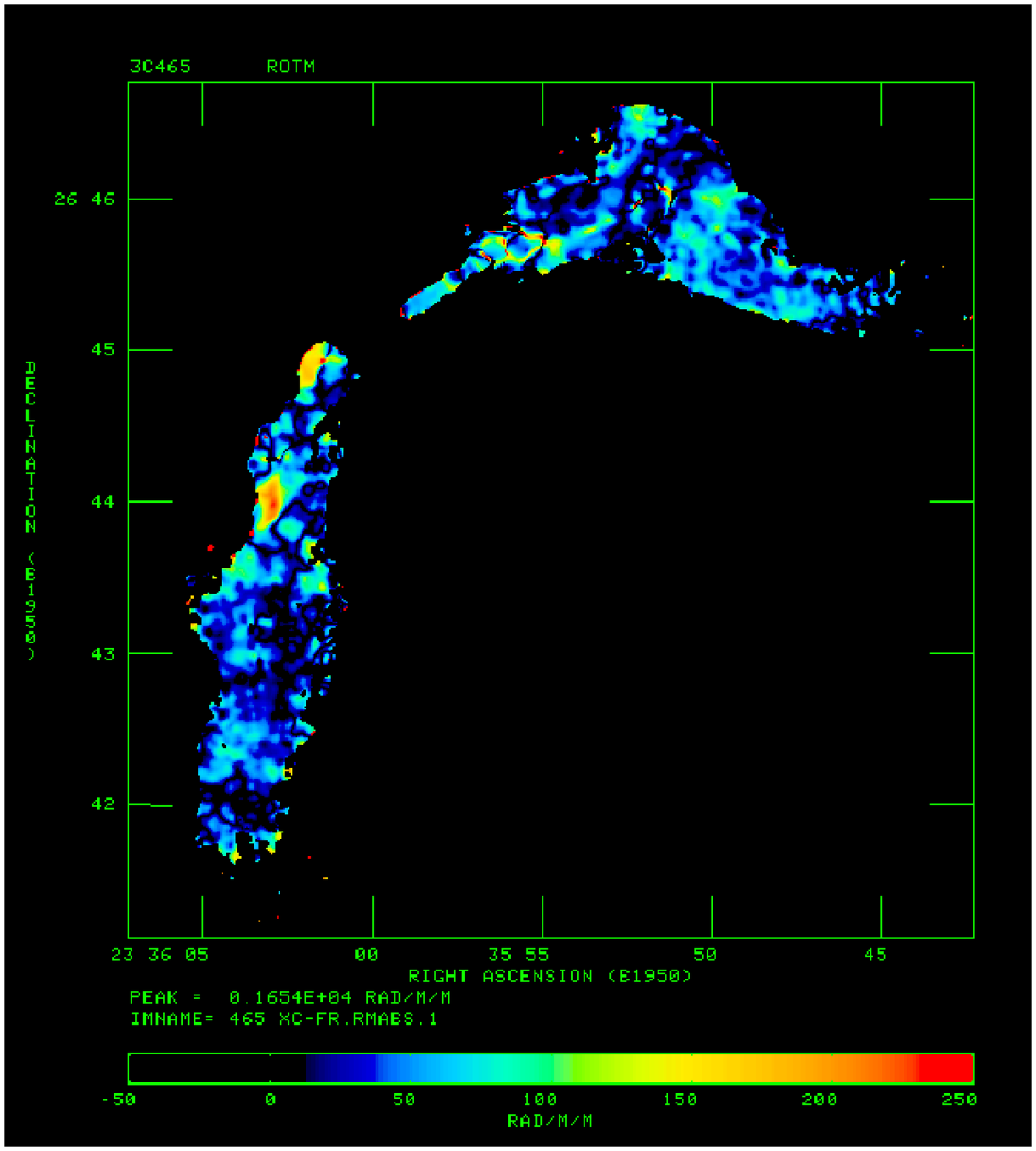}

\figcaption[fig10]{ Absolute value of the RM in 3C75.  The colors are
chosen to highlight extreme values of the RM:  the bright yellow patches on
the east tail reach -200 \radm\ (northerly patch) and +250  \radm\
(southerly patch). }

\begin{deluxetable}{cccccc}
\tablewidth{0pt}
\tablecaption{\bf X-ray Emitting ICM in A400 and A2634}
\tablehead{
	  \multicolumn{3}{c}{Abell 400} 
		& \multicolumn{3}{c}{Abell 2634} 
\\
\colhead{$\sqrt{ab}^1$ (kpc)} & \colhead{$n$(cm$^{-3}$)}  &\colhead
	  { $T$ (keV)}
	& \colhead{$\sqrt{ab}^1$ (kpc)}	& \colhead{$n$ (cm$^{-3}$)}
        &\colhead{ $T$ (keV)}
							}
\startdata
10  & $ 6.5 \times 10^{-3}$ & 1.0  & \nodata & \nodata & \nodata
\\
25 & $ 3.0 \times 10^{-3} $ & 1.0 & 25  & $4.7 \times 10^{-3}$
	   & 1.0 
\\
50 & $ 2.1 \times 10^{-3} $ & 1.5 & 50 & $1.6 \times 10^{-3} $ & 1.0
\\
90 & $ 1.4 \times 10^{-3} $ & 1.5 & 80 & $1.1 \times 10^{-3} $ & 2.2
\\
175 & $ 8.4 \times 10^{-4} $ & 1.4 & 150 & $8.3 \times 10^{-4} $ & 3.0
\\
330 & $ 3.5 \times 10^{-4} $ & 1.4 & 290 & $4.8 \times 10^{-4} $ & 2.4
\\
530 & $ 1.5 \times 10^{-4} $ & 1.6 & 505 & $2.7 \times 10^{-4} $ & 2.4
\\
750 & $ 1.0 \times 10^{-4} $ & 1.6 & 780 & $6.2 \times 10^{-4} $ & 1.4
\\

\enddata
\tablenotetext{1}{$\sqrt{ab}$ is the geometric mean of the major and
minor semi-axes, based on ellipse fits to the surface brightness.}
\end{deluxetable}

\begin{deluxetable}{lcccccc}
\tablewidth{0pt}
\tablecaption{\bf Magnetic Field Models in A400 and A2634}
\tablehead{
	\colhead{Model} & \multicolumn{3}{c}{Abell 400} 
		& \multicolumn{3}{c}{Abell 2634} 
\\
\colhead{} &\colhead{$L_{\rm RM}^1$ (kpc)}
		 & \colhead{$\Bparmean$ ($\mu$G)}  &\colhead{ $p_B / p_g$}
          &\colhead{$L_{\rm RM}^1$ (kpc)}
			 &\colhead{$\Bparmean$ ($\mu$G)}  & \colhead{$p_B / p_g$} 
							}	
\startdata

thin skin  &  1.2  & 25  & 14. & 0.7   & 96 & 420.
\\
single feature     & 10. & 2.9 & .71 & 20. & 3.5 & .52
\\
filled core   & $10\sqrt{10}$  & .93 & .071 & $ 20\sqrt{10}$ & 1.1 & .055
\\
extrema         & 4.4 & 6.7 & 1.1 & 4.1 & 16. & 12.
\\
\enddata
\tablenotetext{1}{$L_{\rm RM}$ is the ``typical'' order scale of the
RM structure, used in equation (2).}
\tablenotetext{2}{Filled core assumes 10 filaments, the maximum possible,
along the line of sight within cluster core.}
\tablenotetext{3}{All models assume $n = .0026$ cm$^{-3}$, $T = 1.5$ keV
for A400;  $n = .0011$ cm$^{-3}$, $T = 1.5$ keV for A2634.}
\end{deluxetable}

\begin{deluxetable}{lcccccccc}
\tablewidth{0pt}
\tablecaption{\bf Cluster Core ICM and Magnetic Fields}
\tablehead{
  	\colhead{Source} & \colhead{$D$}
             &\colhead{$\dot M$} &\colhead{RM}
          &\colhead{$L_{\rm RM}$}  &\colhead{$\langle n \rangle$}
	&\colhead{$T_x$} 
           &\colhead{$\Bparmean^2$}   &\colhead{$p_B / p_g^2$}
\\
     & \colhead{kpc}  & \colhead{$M_{\sun}$/yr}& \colhead{ \radm} 
		&\colhead{kpc}  &\colhead{ cm$^{-3}$}
  & \colhead{keV}  &\colhead{$\mu$G} & 
								}
\startdata
A 400  & 100    &   $\sim 0$ & 50 & 10 & .0021 & 1.5 & 2.9  &  .19
\\
A 1795  &  7    & 400   & 1500 & 3  & .064 & 5.1 & 9.7  &  .021
\\
A 2029  &  8  &  370 & 1500 & 3 & .038 & 7.8 & 16  & .064
\\
A 2052  &  10   &  90  & 400  & 2  & .022 & 2.4 & 11 & 0.17
\\
A 2199  &  30   &  150  & 750 & 3  & .020 & 2 & 15  &  .44
\\
A 2634 &   140  &    $\sim 0$ & 65 & 20 & .0011 & 1.5 & 3.5 & .55
\\
A 3526 & 3 & 30 & 500 & 1 & 0.10 & 3 & 6.5 & .01 
\\
A 4059  &  10   &   120  & 2000 & 3 & .016 & 3.5 & 51 &  3.5
\\
3C129 &  16 & $\sim 0$ & 200 & 3 & .0043 & 6 & 19 & 1.0
\\
Cyg A  &  70    &  200 & 1500 & 10 & .014 & 4 & 15 &  0.22
\\
Hydra A  & 50   & 600 & 2000 & 5 & .0063 & 1.8 & 78 &  40
\\
Virgo A  & 3    &   10  & 2000 & 0.85  & .082 & 1.1 & 35 &   1.0
\\
\enddata
\tablenotetext{1}{The spatial extent over which the RM has been
measured;  may be smaller than the physical extent of the RS.} 
\tablenotetext{2}{All calculations assume one magnetic filament, with
width $L_{RM}$, accounts for the observed RM.  If $N$ filaments
contribute, $\Bparmean$ is reduced by $\sqrt{N}$, and $p_b / p_g$
should be reduced by $N$.}  
\end{deluxetable}
%
%
\begin{deluxetable}{lccccccc}
\tablewidth{0pt}
\tablecaption{\bf Cluster Core Radio Sources and Environment}
\tablehead{
  	\colhead{Source}  &\colhead{$D$ } &\colhead{$B_{mp}$}
          &\colhead{$\Bparmean$}  &\colhead{$P_{\rm rad}$}
           &\colhead{$p_g$}   &\colhead{$\tau_E$}
\\
      & \colhead{kpc}& \colhead{$\mu$G} 
		&\colhead{$\mu$G}  &\colhead{erg/s}  
		   &\colhead{dyn/cm$^2$} &\colhead{Myr} 
								}
\startdata
%
A 400             &  100 & 7 & 2.9 & 
	$7.4 \times 10^{41}$ & $5.1 \times 10^{-12}$ & 240 
\\
A 1795            & 7 & 50 &9.7 & 
	$1.6 \times 10^{41}$& $ 5.2 \times 10^{-10}$ & 39
\\
A 2029            & 8 & 22 & 16 & 
	$3.1 \times 10^{42} $& $4.7 \times 10^{-10}$ & 2.7  
\\
A 2052            & 10 & 25 & 11 & 
	$1.6 \times 10^{42}$& $9.0 \times 10^{-11}$ & 1.8 
\\
A 2199            & 30  & 15 & 15 & 
	$1.9 \times 10^{42}$ & $6.4 \times 10^{-11}$ & 32 
\\
A 2634           & 140 & 6 & 3.5 & 
	$1.1 \times 10^{42}$ & $2.6 \times 10^{-12}$ & 230
\\
A 3526$^1$     &  3  &  \nodata  & 6.5 &
        $ 7.4 \times 10^{40}$ & $ 4.8 \times 10^{-10}$& 13
\\
A 4059            &  10 & 40 & 51 & 
	$2.0 \times 10^{41}$ & $9.0 \times 10^{-11}$ & 16
\\
3C129$^1$      &  16  &  \nodata   & 19 &
        $ 1.5 \times 10^{41} $ & $ 4.1 \times 10^{-11}$& 82
\\ 
Cyg A            & 70 & 25 & 15 & 
	$2.6 \times 10^{44}$ & $9.0 \times 10^{-11}$ & 4.1 
\\
Hydra A          &  50 & 27 & 18 & 
	$1.8 \times 10^{43}$ & $1.8 \times 10^{-11}$ & 4.4  
\\ 
Virgo A             &  3 & 22 & 35 & 
	$5.4 \times 10^{41}$ & $1.4 \times 10^{-10}$ &  0.25 
\\ 
\enddata
\tablenotetext{1}{$B_{mp}$ data not available}

\end{deluxetable}

\end{document}